\documentclass[twocolumn,showpacs,preprintnumbers,amsmath,amssymb]{revtex4}

\usepackage{graphicx}
\usepackage{dcolumn}
\usepackage{bm}

\begin{document}


\title{Alpha-decay properties of superheavy elements $Z=113-125$
in the relativistic mean-field theory with vector self-coupling of
$\omega$ meson}

\author{M.M. Sharma and A.R. Farhan}
 \affiliation{Physics Department, Kuwait University, Kuwait 13060}

\author{G. M\"unzenberg}
\affiliation{Gesellschaft f\"ur Schwerionenforschung, Planckstrasse 1,
D-64220, Darmstadt, Germany}

\date{\today}

\begin{abstract}
We have investigated properties of $\alpha$-decay chains of
recently produced superheavy elements $Z=115$ and $Z=113$ using
the new Lagrangian model NL-SV1 with inclusion of the
vector self-coupling of $\omega$ meson in the framework of the
relativistic mean-field theory. It is shown that the experimentally
observed alpha-decay energies and half-lives are reproduced well
by this Lagrangian model. Further calculations for the heavier
elements with Z=117-125 show that these nuclei are superdeformed
with a prolate shape in the ground state. A superdeformed
shell-closure at $Z=118$ lends an additional binding and an extra
stability to nuclei in this region. Consequently, it is predicted
that  the corresponding $Q_\alpha$ values provide $\alpha$-decay
half-lives for heavier superheavy nuclei within the experimentally
feasible conditions. The results are compared with those of
macroscopic-microscopic approaches. A perspective of the
difference in shell effects amongst various approaches is
presented and its consequences on superheavy nuclei are discussed.
\end{abstract}

\maketitle

\section{Introduction}
\label{intro}
The large progress in superheavy element research has delivered
challenging experimental data.  This provides nuclear structure theory 
an opportunity to interpret the new experimental results and to
improvise on predictions for the location of the superheavy nuclei on
the basis of the new experimental data. At present there exist two 
sets of data due to two methods employed for superheavy-element
synthesis\cite{Hofmann.00}.

The heaviest elements known today, the elements with $Z=$ 110, 111,
and 112, have been produced by cold fusion of heavy ions. Targets
of $^{208}$Pb or $^{209}$Bi were irradiated with the appropriate
projectiles $^{64}$Ni or $^{70}$Zn to produce the isotopes
$^{271}$110, $^{272}$111, and $^{277}$112 \cite{Hofmann.96,Hofmann.02}. 
These have been identified by long $\alpha$ decay chains leading 
to known isotopes. These data are on a safer ground, for they have 
been verified in different laboratories\cite{Ginter.03,Morita.03}. 
Recently, even the heaviest species in this series, $^{277}112$, has 
been confirmed independently \cite{Hofmann.02,Morita.04}. The important 
new result of these experiments is the discovery of a region of deformed,
shell stabilized nuclei \cite{Muenzenberg.88}, centered at $Z=108$
(Hassium) and $N=162$. Theoretical calculations have argued that these
nuclei are stabilized due to a hexadecapole deformation in the
ground state\cite{Cwiok.83,Moeller.95,Boening.86}.

The second approach to synthesize heavy elements leads to elements with
$Z=$ 114, 115, and 116. Beams of $^{48}$Ca on actinide targets e.g.
$^{244}$Pu and $^{245,248}$Cm were used. The isotopes of $^{286-289}$114, 
and $^{290-293}$116 have been identified in various
experiments \cite{Oganessian.99,Oganessian.00,Oganess.00,Oganessian.04}.
Recently, the isotopes $^{287,288}$115 have been observed in
irradiations of $^{243}$Am target with $^{48}$Ca beam \cite{Oganess.04}. 
These results are exciting and of special attraction for theory as the
new region is located already close to the expected superheavy
nuclei. The experimental problem is that the observed nuclides
decay over long $\alpha$-decay chains ending in spontaneous
fission. These form an island of nuclei in itself and
cannot be connected to the known region of isotopes. This makes an
unambiguous identification with the presently used parent-daughter
method impossible. To cope with this problem, a number of
consistency checks have been made \cite{Oganessian.04}. In addition,
first promising experiments to verify the chemical nature of these
elements are underway \cite{Tuerler.04}.

Experimental investigations of superheavy elements encounter a
general problem that only a few atomic nuclei of each species 
are produced. Presently, for the heaviest species only basic properties 
can be directly extracted from experiment. These are 
$Q_{\alpha}$-values and lifetimes. Most of the known nuclides 
have odd proton or neutron numbers and are not likely to decay 
by ground state transitions. Moreover, superheavy nuclei have a 
complicated level structure with high spin and low spin states 
being close together. They are likely to form isomers \cite{Hofmann.00}. 
This makes a comparison with theoretical predictions a little difficult. 
In those cases where the chains end in fission, $Q_{\alpha}$-values and 
lifetimes are the only experimental information that can be extracted 
for nuclides along the chain.  However, ground state masses of nuclei
cannot be obtained.

Properties and structure of superheavy nuclei have been 
investigated extensively using various approaches. 
The approaches consist of microscopic nature such as 
non-relativistic density-dependent Skyrme Hartree-Fock (SHF) 
theory and the Relativistic Mean-Field (RMF) theory or 
macroscopic-microscopic type. In the latter category,
total binding energy of nuclei is obtained as a sum of a smooth
energy based upon liquid drop type formula on which shell
correction is imposed using the method of Strutinsky
\cite{Stru.67,Stru.68}. The most notable effort in this direction
has been the Finite-Range Droplet Model (FRDM) \cite{FRDM.95}. The
shell correction energies were calculated in the FRDM in order to
identify major magic numbers in the region of
superheavy nuclei. The FRDM predicts a major proton magic number
at $Z=114$ beyond the well known magic number $Z=82$. Experimental
data, however, give little support to this magic number.
Calculations using the macroscopic-microscopic approach have also
been performed by another group \cite{Sobicz.01,Muntian.01,
Muntian.03,Muntian.03a}, where Yukawa-plus-exponential model has
been used for the macroscopic component and the Strutinsky shell
correction is used for the microscopic component. Calculations of
properties of superheavy nuclei have been rather successful in the
macroscopic-microscopic approaches.

Extensive studies of superheavy nuclei have also been performed within
self-consistent mean-field models of SHF type \cite{Cwiok.96} or
comparative studies of SHF models and RMF models
\cite{Rutz.97,Bender.99,Bender.01,Rein.02,Bender.03} have been done. 
One of the first studies of the properties of superheavy nuclei within
the RMF theory was performed earlier in ref. \cite{Lala.96} using
the force NL-SH \cite{SNR.93}. It was shown that neutron 
number $N=184$ appears to be magic for lighter superheavy nuclei,
whereas it diminishes for heavier superheavies for $Z>114$. In
this work shell correction energies were also calculated for
heaviest deformed superheavy nuclei. For the proton number, there
was an indication of a deformed shell closure at $Z=114$. This was
consistent with the predictions of the FRDM for a
strong shell gap at $Z=114$ in the deformed region.

In order to identify magic numbers in the region of superheavy elements
theoretically, extensive shell correction calculations have been performed
by Kruppa et al. \cite{Kruppa.00} for a set of the Skyrme and RMF
forces. It is shown that Skyrme models predict the strongest shell
effects at $N=184$ and $Z=124, 126$, but not at $Z=114$. On the other
hand, several RMF forces considered in this work \cite{Kruppa.00} do
not show a shell gap $N=184$. This contrasting difference in the
behavior of magic numbers between the Skyrme forces and RMF forces
can be attributed to the difference in the shell structure in the RMF 
and SHF approaches. This derives partly from the difference 
in the spin-orbit interaction and its isospin dependence.
It was shown in ref. \cite{SLK.94} that isospin dependence in the
Skyrme Hartree-Fock theory is different than that in the RMF theory.
This difference in the isospin dependence of the spin-orbit
potential was shown to be responsible for a correct description of
anomalous isotope shifts in Pb isotopes \cite{SLK.94}. A modification
in the isospin dependence of the spin-orbit potential \cite{SLK.94,Rein.95}
in the Skyrme Ansatz allowed a proper description of the isotope shifts in
Pb nuclei, though refs. \cite{SLK.94} and \cite{Rein.95} provide for
different prescriptions for the dependence. It is, therefore, expected
that Skyrme forces without such a modification would yield results
different than those from the RMF forces. 

A significant difference between the RMF theory and SHF theory 
has emerged in a systematic study of fission barriers of actinide 
nuclei. A comparative study of fission barriers has been
performed recently in SHF models and RMF theory \cite{buervenich.04}. 
It has been shown that the RMF forces considered in this work predict 
lower fission barriers than most of the SHF models. It is, however, not 
clear whether this discrepancy holds for all the available RMF forces. 
This difference between RMF 
and SHF forces, if it holds in general, needs to be understood. 
It is surmised that this difference may lie in the difference in 
shell structure between the two theories.

We have studied the recent $\alpha$-decay chains of 
$Z=115$ \cite{Oganess.04} employing the RMF theory in the present paper.
As it is well established, the Relativistic Mean-Field (RMF) theory
\cite{SW.86} has proved to be successful in providing a framework for
description of various facets of nuclear properties
\cite{Rein.89,GRT.90,Ser.92,SNR.93,Ring.96,SW.98}. The relativistic
Lorentz covariance of the theory allows an intrinsic
spin-orbit interaction based upon exchange of $\sigma$- and
$\omega$-mesons. This has been shown to be advantageous for 
properties such as anomalous isotope shifts in 
Pb nuclei \cite{SLR.93}. The RMF theory has achieved significant 
success in respect of nuclei near the stability
line as well as for nuclei far away from the stability line
\cite{GRT.90,SNR.93,Ring.96}. The nonlinear scalar self-coupling
of $\sigma$-meson  has been the most successful model in the RMF theory
used so far. However, a closer analysis of Lagrangians with the
non-linear scalar coupling of $\sigma$-meson has shown
\cite{Sharma.00} that shell effects with these model Lagrangians
at the stability line are stronger than the experimental data.
This effect is extended to regions far beyond the stability line.
It has been shown that the nature of the shell effects along the
stability line does in turn influence the shell effects
in the extreme and unknown regions \cite{Sharma.02}.

In the present paper, we have calculated properties of $\alpha$-decay
chains of the newly discovered superheavy nuclei $^{288}$115 and
$^{287}$115 \cite{Oganess.04} using the Lagrangian model of the vector
self-coupling of $\omega$ meson. In Section II, we present a perspective
of the shell effects with a view to show a relationship of the shell
effects in the r-process region to those in the superheavy region.
We allude briefly to some of the distinctive features of the RMF theory
in Section II. We will also refer to some of these features in our 
discussion. A brief formalism of the RMF theory is presented in 
Section III, and details of RMF calculations are provided 
in Section IV. Section V discusses results of the RMF calculations 
on $Q_\alpha$ values,
$\alpha$-decay half-lives and deformation properties. A comparison
of the results on $Q_\alpha$ values and $\alpha$-decay half-lives
is made with some earlier calculations and also with predictions
made by some macroscopic-microscopic models. We also predict
properties of heavier superheavy nuclei which can be $\alpha$-decay
precursors of the nuclei $^{288}$115 and $^{287}$115. Deformed
single-particle spectra are presented in order to visualize
existence of possible islands of stability in the deformed region. 
The last section summarizes the results in view of experimental 
feasibility of synthesis of heavier superheavy nuclei.

\section{Importance of the shell effects}

The spin-orbit interaction and consequently how the shell effects
behave in the extreme regions plays a significant role in carving
out shell gaps in superheavy nuclei and in nuclei near r-process
path. In this respect, the RMF theory has an inherent advantage 
in that the spin-orbit interaction arises naturally 
as a result of the Lorentz-Dirac structure of nucleons. 
Consequently, the RMF theory has shown an immense potential 
in being able to describe properties of nuclei along the 
stability line and also for a large number of nuclei beyond 
the stability line. 

Recently, it was shown \cite{Sharma.00} that Lagrangian models with 
the non-linear scalar coupling of $\sigma$ meson only overestimate 
the shell effects at the stability line. In order to remedy the 
problem of strong shell effects, the nonlinear vector self-coupling 
of $\omega$ meson was introduced in Ref. \cite{Sharma.00}.
It was shown \cite{Sharma.00} that the ensuing Lagrangian parameter 
set NL-SV1 is able to reprodduce the shell effects in Ni and Sn isotopes
near the stability line. It was also shown \cite{Sharma.02} that 
the force NL-SV1 based upon the vector self-coupling of $\omega$-meson 
is also successful in describing the available data on the shell effects
across the waiting-point nucleus $^{80}$Zn. This exemplifies the 
importance of an appropriate description of the shell effects along the
line of stability with a view to be able to extrapolate the shell effects
in the extreme regions such as near the r-process path. Due to this
reason, we consider that a description of superheavy nuclei in the 
extreme regions can be put on a footing similar to that of nuclei 
which are close to the r-process path.

\section{Relativistic Mean-Field Theory\protect}

The RMF approach \cite{SW.86} is based upon the Lagrangian density
which consists of fields due to the various mesons interacting with
nucleons. The mesons include the isoscalar scalar $\sigma$-meson, the 
isovector vector $\omega$-meson and the isovector vector $\rho$-meson.
The Lagrangian density is given by:
\begin{eqnarray}
{\cal L}&=& \bar\psi \left( \rlap{/}p - g_\omega\rlap{/}\omega -
g_\rho\rlap{/}\vec\rho\vec\tau - \frac{1}{2}e(1 - \tau_3)\rlap{\,/}A -
g_\sigma\sigma - M\right)\psi\nonumber\\
&&+\frac{1}{2}\partial_\mu\sigma\partial^\mu\sigma-U(\sigma)
-\frac{1}{4}\Omega_{\mu\nu}\Omega^{\mu\nu}+ \frac{1}{2}
m^2_\omega\omega_\mu\omega^\mu\\ &&+\frac{1}{2}g_4(\omega_\mu\omega^\mu)^2
-\frac{1}{4}\bm {R}_{\mu\nu}\bm {R}^{\mu\nu}+
\frac{1}{2} m^2_\rho\bm{\rho}_\mu\bm{\rho}^\mu -\frac{1}{4}F_{\mu\nu}F^{\mu\nu}
\nonumber
\end{eqnarray}
The bold-faced letters indicate the vector quantities.
Here M, m$_{\sigma}$, m$_{\omega}$ and m$_{\rho}$ denote the nucleon-,
the $\sigma$-, the $\omega$- and the $\rho$-meson masses respectively,
whereas g$_{\sigma}$, g$_{\omega}$, g$_{\rho}$ and e$^2$/4$\pi$ = 1/137 are
the corresponding coupling constants for the mesons and the photon,
respectively.

The nonlinear scalar interaction has been used extensively to describe
the ground-state properties of finite nuclei. Herein, the $\sigma$ meson
is assumed to move in a scalar potential of the form:
\begin{equation}
U(\sigma)~=\frac{1}{2} m_{\sigma}^{2} \sigma^{2}~+~
\frac{1}{3} g_{2}\sigma^{3}~+~\frac{1}{4} g_{3}\sigma^{4}.
\end{equation}
where $g_2$ and $g_3$ represent the parameters of the nonlinear coupling
of $\sigma$ meson to nucleon. This model Lagrangian has been
very successful in describing properties of nuclei at the stability 
line and also for nuclei away from the stability line.The coupling 
constant for the non-linear $\omega$-term is denoted by $g_4$ in the 
Lagrangian of Eq. (1). The vector self-coupling of the $\omega$-meson 
was first proposed in ref. \cite{Bod.91}, where properties of nuclear matter
associated to this potential were discussed. It is interesting to
note that introduction of the non-linear coupling of $\omega$-meson
softens the equation of state (EOS) of the nuclear matter significantly.
This has the consequence that the maximum neutron star mass with such
an EOS appears within the bounds of empirically observed values.

\section{Details of the calculations\protect}

The RMF calculations in this work have been performed in an axially
symmetric deformed basis.  The method of expansion \cite{GRT.90} of 
wavefunctions into harmonic-oscillator basis has been used to solve 
the Dirac and Klein-Gordan equations. Both the fermionic and bosonic 
fields have been expanded in a harmonic oscillator basis of 20 shells. 
The pairing has been included within the BCS scheme employing constant 
pairing gaps. The prescription of ref. \cite{MN.92} which provides for 
a best fit of pairing gaps for neutrons and protons over a large
range of nuclei has been used. Accordingly,  pairing gaps
are taken as
\begin{equation}
\Delta_{n(p)}~= 4.8N^{-1/3}(Z^{-1/3})~
\end{equation}
where $N$ and $Z$ represent the neutron and proton number of a nucleus.
The Lagrangian parameter set NL-SV1 \cite{Sharma.00} has been used
for all the calculations. The parameters of the force NL-SV1 are
presented in Table I. In this work, we have not included the
blocking of pairing for an odd particle. Extensive calculations
with the blocking of pairing would be presented elsewhere.

\begin{table}
\caption{The Lagrangian parameters of the force NL-SV1 used in
the calculations. All the masses are in MeV. While $g_2$ is in $fm^{-1}$,
all the other coupling constants are dimensionless. Here $g_4$ represents
the quartic coupling of $\omega$ meson.}
\begin{ruledtabular}
\begin{tabular}{lll}
M            =  939.0   & $g_{\sigma}$ =10.1248 & $g_{2}$ = -9.2406 \\
$m_{\sigma}$=510.0349    &  $g_{\omega}$   =12.7266 & $g_{3}$= -15.388 \\
$m_{\omega}$=783.0      &  $g_{\rho}$     =  4.4920 &  $g_{4}$= 41.0102\\
$m_{\rho}$=763.0        &                           &  \\
\end{tabular}
\end{ruledtabular}
\end{table}

We have carried out constrained calculations with a quadratic
constraint. The potential energy landscapes of nuclei over a
large range of quadrupole deformation have been mapped out
and various possible minima in the total energy have been
identified. This also serves to confirm the total binding energy
of ground state as obtained in individual RMF+BCS minimizations.

\section{Results and Discussion}

\subsection{$Q_\alpha$ values and half-lives}

The decay chain of the superheavy nucleus $^{288}$115 (Z=115) was
observed \cite{Oganess.04} for 3 events in the
3n evaporation channel in the reaction $^{243}$Am + $^{48}$Ca with the
projectile energy $E = 248$ MeV. The energy of alpha particles emitted
as a result of consecutive $\alpha$-decay of various product nuclei
is within about 250 keV in the three events. The maximum dispersion in
the alpha energy has been observed for $^{284}$113 amounting to about
0.5 MeV. The best two values agreeing with each other
have been selected as the acceptable value for $Q_\alpha$ in
ref. \cite{Oganess.04}.

\begin{figure}
\centering
\resizebox{0.60\textwidth}{!}{%
  \rotatebox{270}{\includegraphics{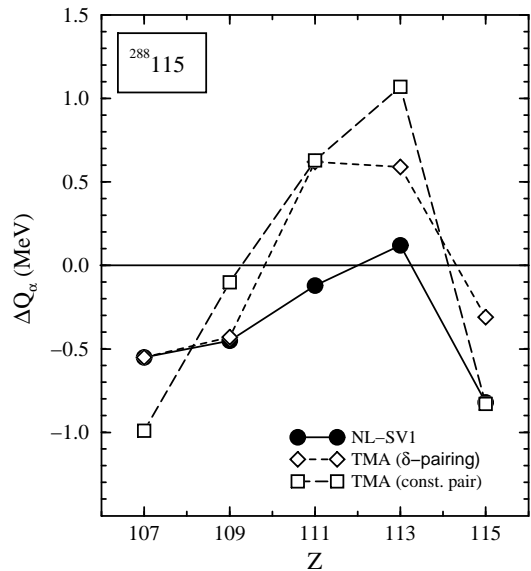}}
}
\caption{The difference $\Delta Q_\alpha$ = $Q_\alpha$ (theory) -
$Q_\alpha$ (expt.) between the theoretical and the experimentally
observed $Q_\alpha$ values for the $\alpha$-decay chain with the parent
nucleus $^{288}$115. The RMF results with NL-SV1 (full circles)
are compared to those with the RMF calculations using the
force TMA with constant pairing gap (open squares) and using a
density-dependent $\delta$-pairing force including blocking 
effect (open diamonds).}
\label{fig:1}       
\end{figure}

\begin{figure}
\resizebox{0.60\textwidth}{!}{%
  \rotatebox{270}{\includegraphics{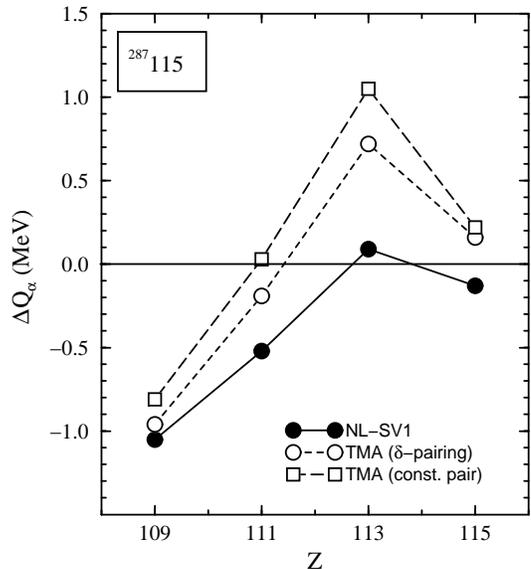}}
}
\caption{The same as for Fig.~\ref{fig:1}, but for the $\alpha$-decay chain
with the parent nucleus $^{287}$115.}
\label{fig:2}       
\end{figure}

Results of the RMF calculations with the Lagrangian set NL-SV1
for the newly discovered chains of superheavy nuclei are shown in 
Figs.~\ref{fig:1} and \ref{fig:2}.
Figure~\ref{fig:1} shows the difference $\Delta Q_\alpha$ of the calculated
$Q_\alpha$ values from the experimentally obtained ones for the chain
with the parent nucleus $^{288}$115. For the nucleus $^{288}$115, 
calculations with NL-SV1 show a deviation from the experimental value
by about 0.8 MeV. For the nuclei $^{284}$113 and $^{280}$111, NL-SV1 shows
a very good agreement with the data. On going to lighter products of
the chain, i.e. for $^{276}$Mt (Z=109)  and $^{272}$Bh (Z=107), our
results agree with the data within about half an MeV.
A comparison is also made with the relativistic mean-field
calculations of ref. \cite{Geng.03} using the parameter set TMA
\cite{Sugahara.94}. Results of TMA with the use of constant pairing
(open square) and a density-dependent $\delta$-pairing force (diamonds)
are shown. The results shown with the $\delta$-pairing also include
the blocking effect. The  $Q_\alpha$ obtained with the constant pairing
for the nucleus $^{288}$115 is about the same as that with NL-SV1. However,
for other nuclei such as $^{284}$113 and $^{272}$Bh (Z=107) the disagreements
with the data amounts to about 1 MeV. Whilst for $^{284}$113, TMA with
constant pairing overestimates the experimental value by about 1 MeV, it
underestimates the experimental $Q_\alpha$ for $^{272}$Bh (Z=107) by about
the same amount. In comparison, the use of the $\delta$-pairing force
with the blocking has shown some improvements in the results with TMA.
It is seen from the figure that our present calculations with NL-SV1
are in good agreement with the  $Q_\alpha$ values of the decay chain
of $^{288}$115 with the exception for the parent nucleus.

In ref. \cite{Oganess.04}, a single decay chain of $^{287}$115 was observed
on increasing the projectile energy by 5 MeV to $E = 253$ MeV. It is
assumed that $4n$ evaporation channel in the excitation function becomes
dominant. It may, however, be said that any other event for this decay chain
has not been observed to confirm the $\alpha$-decay energies. In view of this,
these values may be treated with caution. The $\alpha$-decay energies
for the decay chain of $^{287}$115 as reported in the ref. \cite{Oganess.04}
are 10.59 MeV, 10.12 MeV, 10.37 MeV and 10.33 MeV, respectively. The two
latter values show an increase over that of the second value (10.12 MeV)
in the decay chain. This behaviour is different from that of the $^{287}$115
decay chain, where the $\alpha$-decay energies show a decreasing trend
with the decrease in the Z value of the daughter nuclei. Assuming that
the shell structure has not changed from the $^{288}$115 chain to the
$^{287}$115 chain, there could easily be an uncertainty of about 0.5 MeV
in the $\alpha$-decay energies of $^{279}$111 and $^{275}$109.

\begin{figure}
\resizebox{0.65\textwidth}{!}{%
  \rotatebox{270}{\includegraphics{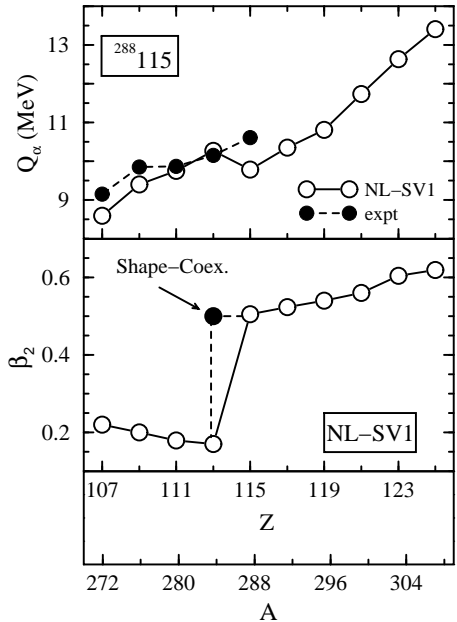}}
}
\caption{The $Q_\alpha$ values obtained with the force NL-SV1
in the RMF calculations are compared with the experimental data from
ref. \cite{Oganess.04} for the $\alpha$-decay chain of $^{288}$115
in the upper panel. The $Q_\alpha$ values predicted by our model for the
heavier parents of this chain are also shown. In the lower panel
the ground-state quadrupole deformation $\beta_2$ are shown  by open
circles. The solid circle for the nucleus $^{284}$113 exhibits a
second minimum at the highly deformed (superdeformed) configuration
that is in coexistence with the low deformed prolate state.}

\label{fig:3}       
\end{figure}

\begin{figure}
\resizebox{0.65\textwidth}{!}{%
  \rotatebox{270}{\includegraphics{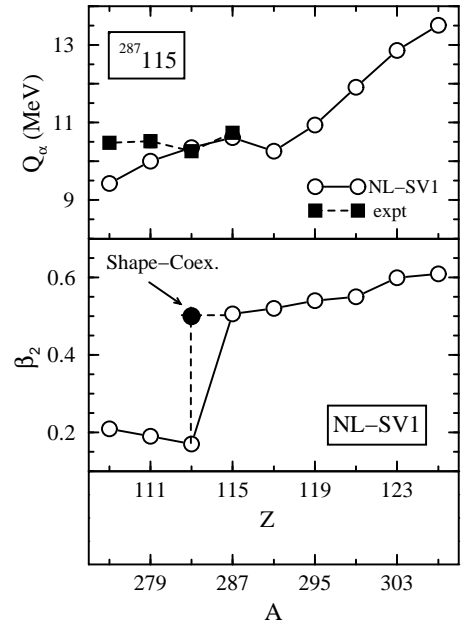}}
}
\caption{The same as for Fig. 3 but for the chain consisting of
the nucleus $^{287}$115.}
\label{fig:4}       
\end{figure}

The results of our calculations for the decay chain of $^{287}$115 are
shown in Fig. 2. The $Q_\alpha$ values with NL-SV1 agree well with
the experimental values for the nuclei $^{287}$115 and  $^{283}$113.
For the nucleus $^{279}$111, our calculations show a $Q_\alpha$ of 10.0
MeV compared to 10.52 MeV as observed in ref. \cite{Oganess.04}.
The difference is accentuated to about 1 MeV for the nucleus $^{275}$109.
It is noted that the calculated values with NL-SV1 show
a decreasing trend with the Z value. An uncertainty in the $\alpha$-decay
energies for this decay chain as commented above may partly account for
this discrepancy. Generally, there is an overall agreement of the NL-SV1
results within 0.5 MeV for the $Q_\alpha$ values save that for 
$^{275}$Mt $(Z=109)$.

Calculations with TMA from ref. \cite{Geng.03} are also compared in
Fig. 2. The calculations both with constant pairing and with $\delta$-pairing
with blocking show a trend similar to that of NL-SV1 vis-a-vis the
experimental values. In the calculations of ref. \cite{Geng.03}
disagreements are more pronounced for $^{283}$113 and $^{275}$Mt
($Z=109$).

\subsection{Predictions for heavier superheavy nuclei}

We have calculated ground-state properties of the decay chains
of ref. \cite{Oganess.04}. We have also extended our calculations
in order to be able to predict properties of odd-Z nuclei heavier than
$^{288}$115 and $^{287}$115, whereby $^{288}$115 and $^{287}$115
would constitute daughter nuclei of a sequential $\alpha$-decay chain.
The $Q_\alpha$ value and quadrupole deformation $\beta_2$ corresponding to
the lowest energy state for the chain comprising the $^{288}$115 and
its daughters are shown in Fig. 3. The NL-SV1 results show general agreement
with the experimental $Q_\alpha$ values \cite{Oganess.04} (see Table II)
as discussed above and shows an increasing trend with an increase in the
$Z$ value. The NL-SV1 value, on the other hand, shows a slight decrease at
$Z=115$. This is indicative of a possible deformed shell closure at
$Z=114$ with NL-SV1. For all the nuclei above $Z=115$, the $Q_\alpha$
value shows a systematic increase with a slight kink at $Z=119$.

\begin{figure}
\resizebox{0.65\textwidth}{!}{%
  \rotatebox{270}{\includegraphics{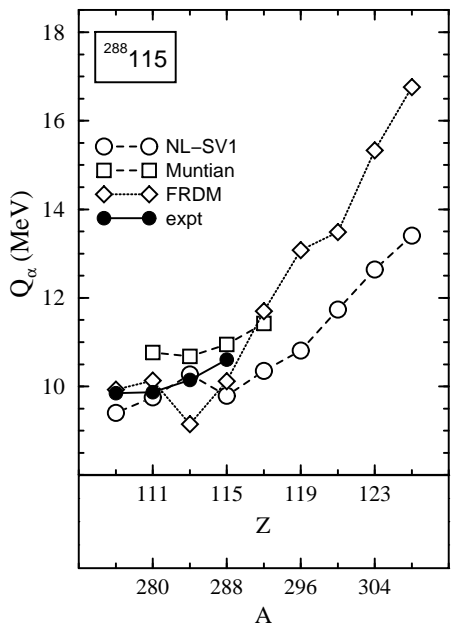}}
}
\caption{A comparison of the $Q_\alpha$ values obtained with
NL-SV1 is made with the results of ref. \cite{Muntian.03} and
FRDM \cite{Moeller.97} for the chain comprising $^{288}$115.
The experimental values obtained from the recent experiment
\cite{Oganess.04} are also shown for comparison.}
\label{fig:5}       
\end{figure}

\begin{figure}
\resizebox{0.65\textwidth}{!}{%
  \rotatebox{270}{\includegraphics{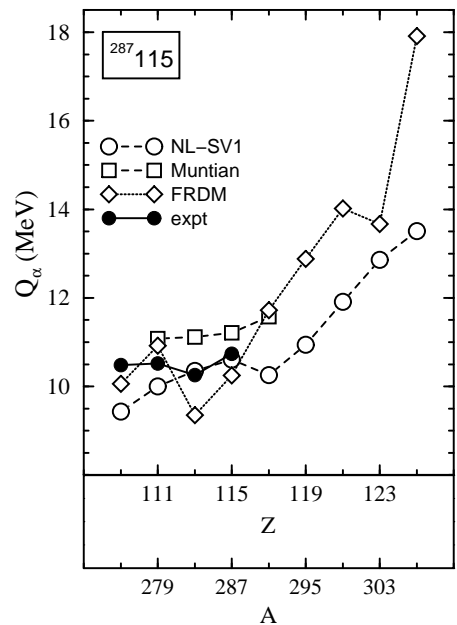}}
}
\caption{The same as for Fig. 5 but for the chain consisting of
the nuclide $^{287}$115.}
\label{fig:6}       
\end{figure}

\begin{table*}
\caption{The $Q_\alpha$ values, $\alpha$-decay half-lives $T_\alpha$,
and quadrupole deformation $\beta_2$ of nuclei in the
decay chain consisting of the nuclide $^{288}$115 as obtained with
RMF calculations using the Lagrangian set NL-SV1. The $Q_\alpha$ values
and estimated half-lives from the recent experimental results
\cite{Oganess.04} are also shown for comparison.}
\begin{ruledtabular}
\begin{tabular}{c c c c c c c c}
&  Nucleus        & $Q_\alpha$ (MeV)   &  $T_\alpha$     &  $\beta_2$ && $Q_\alpha$ (exp.) (MeV) & $T_\alpha$ (exp.) \\
\hline
& $^{308}$125     & 13.41  & 16 $\mu$s & 0.62  &&                  &           \\
& $^{304}$123     & 12.64  & 0.20 ms   & 0.61  &&                  &          \\
& $^{300}$121     & 11.74  & 6.5  ms   & 0.56  &&                  &          \\
& $^{296}$119     & 10.81  & 0.38 s    & 0.54  &&                  &         \\
& $^{292}$117     & 10.35  & 1.63 s    & 0.52  &&                  &         \\
& $^{288}$115     & 9.79   & 15.3 s    & 0.50  && $10.61 \pm 0.06$ & $87^{+105}_{-30}$ ms \\
& $^{284}$113     & 10.27  & 0.15 s    & 0.17  && $10.15 \pm 0.06$ & $0.48^{+0.58}_{-0.17} s $\\
& $^{280}$111     & 9.75   & 0.97 s    & 0.18  && $9.87 \pm 0.06$ & $3.6^{+4.3}_{-1.3}$ s \\
& $^{276}$Mt      & 9.40   & 2.21 s    & 0.20  && $9.85 \pm 0.06$ & $0.72^{+0.87}_{-0.25}$ s \\
& $^{272}$Bh      & 8.60   & 150 s     & 0.26  && $9.15 \pm 0.06$ & $9.8^{+11.7}_{-3.5}$ s  \\
\end{tabular}
\end{ruledtabular}
\end{table*}

The corresponding $\beta_2$ values are shown in the lower panel. The quadrupole
deformation stays in the vicinity of $\beta_2 \sim 0.2 $ in going from
$Z=107$ to $Z=113$. However, the nucleus $^{284}$113 marks a transition point
for the chain, in that it exhibits a shape coexistence of a low prolate
deformation ($\beta_2 = 0.17$) with a high prolate deformation
($\beta_2 = 0.50$) in the  ground state. Here, the state with the
low deformation is the lowest one. All the nuclei including and above
$Z=115$ exhibit a large prolate deformation ($\beta_2 \sim 0.6$) in the ground
state. This is akin to a superdeformation. As the single-particle scheme
would show, this is associated with a strong shell gap at $Z=118$ in the
superdeformed region. This allows an increase in the binding energies
of nuclei and accordingly the $Q_\alpha$ values are comparatively
suppressed. 

The $Q_\alpha$ and $\beta_2$ values for the ground state of nuclei in the
chain associated with $^{287}$115 are shown in Fig. 4. As discussed above,
the $Q_\alpha$ value for the nucleus $^{275}$Mt shows a deviation of about
1 MeV from the observed value. For the others, there is a good agreement
with the experimental value (see Table III). The NL-SV1 values show a
systematic increase in $Q_\alpha$ in going from $Z = 109$ to $Z = 115$.
A kink in $Q_\alpha$ at $Z=115$ indicates a presence of deformed
shell closure at $Z = 114$.

\begin{table*}
\caption{The same as for Table II but for the decay
chain consisting of the nuclide $^{287}$115.}
\begin{ruledtabular}
\begin{tabular}{c c c c c c c c}
&  Nucleus        & $Q_\alpha$ (MeV)   &  $T_\alpha$     &  $\beta_2$ && $Q_\alpha$ (exp.) (MeV) & $T_\alpha$ (exp.) \\
\hline
& $^{307}$125     & 13.51  & 10 $\mu$s   & 0.61  &&                  &           \\
& $^{303}$123     & 12.86  & 68 $\mu$s   & 0.60  &&                  &          \\
& $^{299}$121     & 11.91  & 2.6  ms     & 0.55  &&                  &          \\
& $^{295}$119     & 10.94  & 0.17 s      & 0.54  &&                  &         \\
& $^{291}$117     & 10.26  & 2.9 s       & 0.52  &&                  &         \\
& $^{287}$115     & 10.57  & 100 ms      & 0.51  && $10.74 \pm 0.06$ & $32^{+155}_{-14}$ ms \\
& $^{283}$113     & 10.35  &  90 ms      & 0.17  && $10.26 \pm 0.06$ & $100^{+490}_{-45}$ ms \\
& $^{279}$111     & 10.00  & 200 ms      & 0.19  && $10.52 \pm 0.06$ & $170^{+810}_{-80}$ ms \\
& $^{275}$Mt      & 9.43   & 1.8 s       & 0.21  && $10.48 \pm 0.06$ & $9.7^{+46}_{-4.4}$ ms \\
\end{tabular}
\end{ruledtabular}
\end{table*}

As far as deformation of nuclei in this chain is concerned, all nuclei
with low $Z$ values including $^{283}$113 exhibit a low value of $\beta_2$
in the vicinity of 0.20. Here, the nucleus $^{283}$113 with $Z=113$ exhibits
the phenomenon of shape coexistence of a low prolate deformation
($\beta_2 = 0.17$) with a high prolate deformation ($\beta_2 = 0.50$)
in the  ground state. The state with with the low $\beta_2$ value
is the lowest minimum in the binding energy. This behavior is similar
to that exhibited by the corresponding nucleus $^{284}$113 in the other
chain. Likewise, the nucleus $^{283}$113 marks a transition point in the 
shapes of nuclei in this chain. All the nuclei including and 
above $Z=115$ show a superdeformed ground state. All the 
features shown in Fig. 4 for this chain exhibit a similarity 
to those shown in Fig. 3. Therefore, properties of nuclei for 
both the chains point to almost the same structural behavior.

RMF calculations with another Lagrangian parameter set NL-SV2 
\cite{Sharma.00,Sharma.05} with the vector self-coupling of 
$\omega$-meson for both the chains of Figs. 3 and 4 show that
nuclei $^{284}$113 and $^{283}$113 exhibit a shape coexistence of
low prolate deformation with a high prolate deformation in the 
ground state. The magnitude of deformations with NL-SV2 is seen
to be similar to that obtained with NL-SV1 as shown in Figs.~3 and 4.
With the Lagrangian set NL-SV2, superheavy nuclei heavier than  
$^{284}$113 and $^{283}$113 in these chains are also observed to 
be superdeformed in the ground state. Accordingly, the nuclei 
$^{284}$113 and $^{283}$113 denote a transition point from a low
deformation to a superdeformation also with NL-SV2. Thus, the 
results on the shape-coexistence and the presence of a second 
superdeformed minimum for nuclei with $Z=113$ with NL-SV1 are 
also exhibited by NL-SV2. A comprehensive comparison of results 
in the superheavy region wtih various Lagragian models will be 
presented elsewhere \cite{Farhan.05}.

In a recent investigation of deformation properties of superheavy nuclei
\cite{Muntian.04}, it has been noted that macroscopic-microscopic approach
does not support a superdeformation in the superheavy region. This difference
in the outcome of the results of ref. \cite{Muntian.04} can be understood
due to strong shell effects prevalent in the macroscopic-microscopic
approaches. The strong shell effects or equivalently strong shell gaps
would disfavour a superdeformation.

Our results for the recently discovered chain of $^{288}$115 and 
predictions for heavier superheavy nuclei are compared with some
other models in Fig. 5. Detailed macroscopic-microscopic calculations
for a large number of nuclei in the superheavy region of $Z=110-120$
have recently been performed by Muntian et al. \cite{Muntian.03}.
The results available on some nuclei in this chain are shown.
Whilst this model shows a reasonably good
agreement with the experimental values on $^{288}$115 and $^{284}$113, it
overestimates the datum for $^{280}$111. It is interesting to compare
the data also with those from the macroscopic-microscopic
model FRDM \cite{Moeller.97}. The FRDM values show an agreement with the 
data for two lighter superheavy nuclei and also for $^{288}$115 within 
about  0.5 MeV. However, for the nucleus $^{284}$113 the FRDM value 
underestimates the $Q_\alpha$ value by $\sim$ 1 MeV. For heavier
$\alpha$-decay precursors of $^{288}$115, i.e., nuclei with $Z=117-125$, 
we can compare our results with NL-SV1 with those from FRDM.

As far as a comparison of the macroscopic-microscopic calculations of
Muntian et al. \cite{Muntian.03} with FRDM are concerned,
the predictions only for $^{292}$117 can be compared. For this
nucleus the two models agree. As further results
on other odd nuclei from ref. \cite{Muntian.03} are not available,
it is difficult to compare the two models in the context of the two
chains studied in our work. However, as predictions from
ref. \cite{Muntian.03} are generally on the higher side in 
this region, we surmise that the results of the two 
macroscopic-microscopic would follow each other for heavier
superheavies. In comparison, our results with NL-SV1 show a smooth
increase in the value of $Q_\alpha$ with $Z$ value. The NL-SV1 results,
on the other hand, predict $Q_\alpha$ values which are smaller by about
1-2 MeV from those of the FRDM. This is due to an extra stability
lent by the superdeformed ground state of all the heavier nuclei
with $Z \ge 115$. This will be demonstrated by a large shell gap in
the deformed single-particle spectrum at $Z=118$ and $Z=120$.

Calculation of half-lives of $\alpha$-decay from $Q_\alpha$ values 
is not free of uncertainties. This is subject to Viola-Seaborg
systematics fitted over a given region of nuclei. The mostly used
systematics derive from a fit of parameters done in ref. \cite{Sobiz.89}.
We have calculated the $\alpha$-decay half-lives of nuclei using the
formula from Viola-Seaborg systematics. Accordingly, the $\alpha$-decay
half-life can be written as
\begin{equation}
log~T_\alpha (s) = (aZ + b)Q_\alpha^{-1/2} + (cZ + d)~
\end{equation}
where $Z$ is proton number and $Q_\alpha$ is alpha-decay energy of the 
parent nucleus in MeV. The parameters are $a = 1.66175$, $b = -8.5166$,
$c= -0.20228$ and $d= -33.9069$. These parameters are taken from
ref. \cite{Sobiz.89}, where these were readjusted in order to take into
account new data. These parameters have been found to be successful
over a broad range of nuclei.

The $T_\alpha$ values calculated using $Q_\alpha$ values from
NL-SV1 for nuclei of the chain consisting of $^{288}$115 are
shown in Table II. The experimentally observed
$\alpha$-decay half-life \cite{Oganess.04} of $^{288}$115 and its
daughters are also shown for comparison. It may be noted that experimental
values of $T_\alpha$ for this chain are based upon 3 events observed in
the experiment \cite{Oganess.04}. The corresponding half-life quoted in
ref. \cite{Oganess.04} has been obtained from the average of the life-times
observed in the experiment. Consequently, one can notice large error
bars in the experimental values. These values therefore signify the
order of magnitude of a $T_\alpha$ value.

Given the large uncertainties in the experimental values,
$T_\alpha$ values calculated with NL-SV1 agree well with the
experimental ones with the exception for $^{288}$115 and $^{272}$Bh
(see Table II). The theoretical half-life for these two nuclei are about
an order of magnitude larger than the experimental values. This is
due to the reason that with our model the $Q_\alpha$ values are
$\sim 0.50-0.80$ MeV smaller than the experimental $Q_\alpha$ value.
For other superheavy nuclei with $Z=109-113$, there is a good
agreement between the theoretical and experimental $\alpha$-decay
half-lives. This is due to a good agreement between the corresponding
$Q_\alpha$ values.

In going to the heavier precursor nuclei in the same chain (Table II),
there is an increasing tendency in the value of $Q_\alpha$ with an increase
in the $Z$ value. However,  $Q_\alpha$ values with NL-SV1 are systematically
lower than those of FRDM. The lower $Q_\alpha$ values with NL-SV1 has
the consequence that superheavy nuclei with high $Z$ such as $Z=117$,
$Z=119$ and $Z=121$ would not have a very small $\alpha$-decay half-life
as compared to that from FRDM. The $\alpha$-decay half-lives predicted from
NL-SV1 for the nuclei $^{292}$117, $^{296}$119, and $^{300}$121 are
$\sim 1.6~s,~0.4~s$ and $7~ms$, respectively. This would put these nuclei
within the range of experimental accessibility and feasibility.

A comparison of various results for the chain comprising $^{287}$115
are shown in Fig. 6. In general, the results of Muntian et al.
\cite{Muntian.03} overestimate the new data by $\sim$ 0.5-1.0 MeV. On the
other hand, the results of FRDM for this chain are very similar to those
for the other chain as shown in Fig. 5. Though the FRDM results show a
good agreement with a few data, kinks at $Z=111$ and $Z=113$ show
an undulating character of the FRDM results. Such kinks are not
present in other results. This may be due to a strong spherical shell
closure at $Z=114$ as predicted by the FRDM.

The $\alpha$-decay half-lives for the decay chain of $^{287}$115
from NL-SV1 are compared with the experimental estimates from
ref. \cite{Oganess.04} in Table III. It may be noted that the experimental
value is based upon life-time estimates from a single decay event of the
superheavy nucleus $^{287}$115. There is an overall good agreement between
NL-SV1 values and the experimental ones. Only for $^{275}$Mt ($Z=109$) does
NL-SV1 overestimate the half-life by about 2 orders of magnitude.
The experimental half-life for the nucleus $^{275}$Mt has been estimated
to be $9.7^{+46}_{-4.4} ms$. This value is much lower than the experimental
estimates for the heavier superheavies such as $^{279}$111 and $^{283}$113
(see Table III). In comparison, the experimental $Q_\alpha$ value for
$^{275}$Mt is comparable to that for $^{279}$111. According to Viola-Seaborg
formula, this experimental $Q_\alpha$ for $^{275}$Mt would indicate
$T_\alpha$ value much larger than the quoted one \cite{Oganess.04}.
In view of this, there seems to be a little discrepancy between the
experimentally deduced $Q_\alpha$ and $T_\alpha$ value for the nucleus
$^{275}$Mt. This may be due to ambiguities in identification in a single
event.

The generally higher $Q_\alpha$ values from macroscopic-microscopic
calculations \cite{Muntian.03} vis-a-vis the experimental data
\cite{Oganess.04} by $\sim 0.5-1.0$ MeV would mean a decrease in
$\alpha$-decay half-life by about 1-3 orders of magnitude in the
macroscopic-microscopic calculations \cite{Muntian.03}.  
However, it is worth mentioning that the results of 
ref. \cite{Muntian.03} on $\alpha$-decay energies
describe well the data on and variation in the $Q_\alpha$ values
with neutron numbers for isotopic chains below $Z=109$, as shown
clearly in Fig.~2 of ref.~\cite{Oganess.04}.

For the heavier superheavies of the chain comprising of the nucleus
$^{287}$115 (Fig. 6) the FRDM values show a strong increase in the
$Q_\alpha$ value. This would make the half-life of some of these
nuclei in the range of nanosecond or even smaller. In comparison,
our predictions with NL-SV1 show only a modest increase in the
$Q_\alpha$ values for nuclei heavier than $Z = 115$ (see Table III).
This behaviour is very similar to that exhibited by the other chain
in Fig. 5. The $\alpha$-decay half-life using the Viola-Seaborg systematics
for nuclei $^{292}$117, $^{296}$119, and $^{300}$121 in our model are
2.9 s, 0.17 s and 2.6 ms, respectively. These values are well within
the reach of experimental feasibility. Thus, the NL-SV1 results shown
in Fig. 5 and Fig. 6 puts several heavy elements within the range
of feasibility of realization.

\begin{figure*}
\resizebox{0.65\textwidth}{!}{%
  \rotatebox{270}{\includegraphics{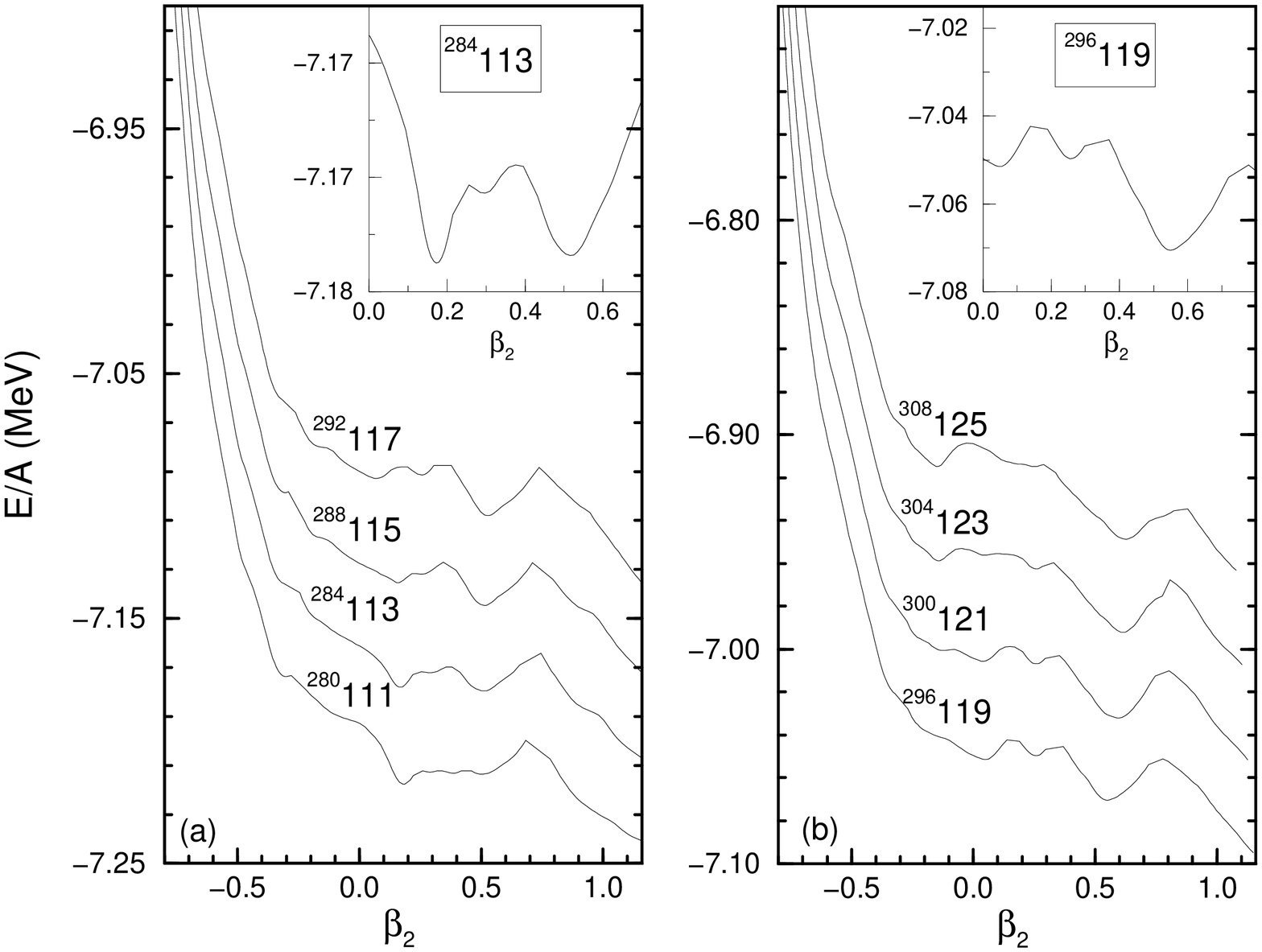}}
}
\caption{Potential energy landscapes of nuclei in the decay chain
comprising $^{288}$115 for (a) lighter superheavies below $Z \le 117$.
The shape coexistence of a deformed and a superdeformed prolate shape
is shown for $^{283}$113 in the inset. (b)Potential energy landscapes of
heavy superheavies with $Z \ge 119$. All the nuclei show a superdeformed
ground state as depicted in the inset for $^{296}$119.}
\label{fig:7}       
\end{figure*}

\begin{figure*}
\resizebox{0.65\textwidth}{!}{%
  \rotatebox{270}{\includegraphics{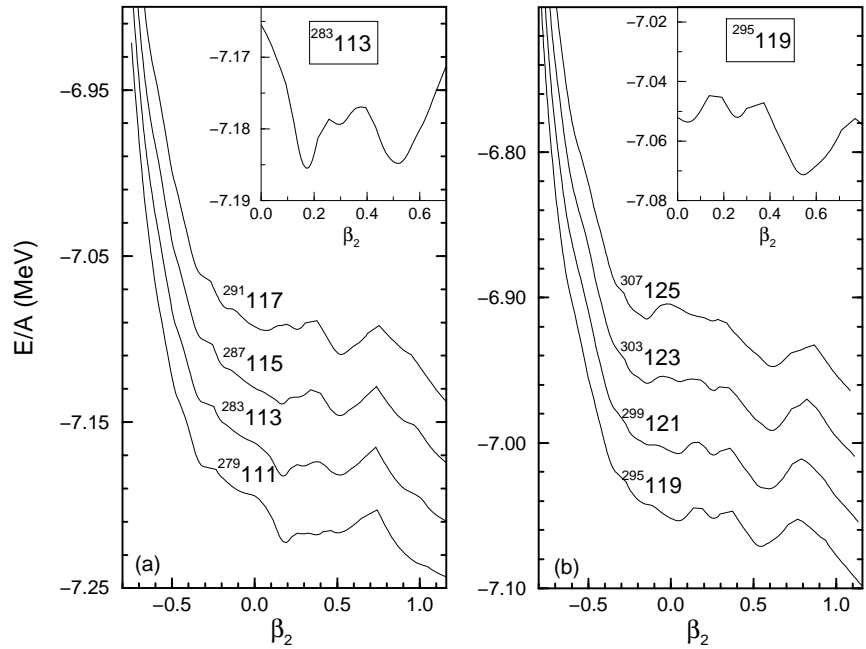}}
}
\caption{The same as for Fig.~7, but for the decay chain comprising
of the nuclide $^{287}$115.}
\label{fig:8}       
\end{figure*}

\subsection{Potential energy landscapes}

We have performed constrained calculations with a quadratic constraint
in order to map out potential energy landscape. The total energy of
nuclei as a function of quadrupole deformation $\beta_2$ is shown in
Figs.~7 and 8 for superheavy nuclei in the two isotopic chains studied
in this work. All the nuclei of chain comprising of both
$^{288}$115 (Fig.~7) and $^{288}$115 (Fig.~8) exhibit a prolate
deformation for all the nuclei from $Z=111$ to $Z=125$.
A configuration in oblate space seems to be forbidden as suggested
by the total potential energy curve for the superheavy nuclei
shown in Figs. 7 and 8.

Consistent with previous description, the nuclei $^{280}$111 and
$^{279}$111 are prolate deformed with a low value of quadrupole
deformation. On the other hand, the nuclei $^{284}$113 (Fig.~7(a))and
$^{283}$113 (Fig.~8(a)) exhibit two minima, as discussed earlier.
The two minima, which coexist in energy are depicted in the insets
of Fig.~7(a) and Fig.~8(a). The second minimum at $\beta_2 \sim 0.50$
lies within a few hundred keV of the first minimum. For all the heavier
elements with $Z > 113$, the ground state minimum is obtained in
the region of a high prolate deformation as seen by a single well in
the curves in part (a) and (b) of Figs.~7 and 8. The high prolate
deformation of superheavies continues in the high $Z$ region
as exemplified in the insets of Fig.~7(b) and Fig.~8(b) for the
element $Z=119$. It is interesting to note that the potential
energy landscapes for Fig.~7 are very similar to those of
Fig.~8.

\begin{figure}
\resizebox{0.50\textwidth}{!}{%
  \rotatebox{0}{\includegraphics{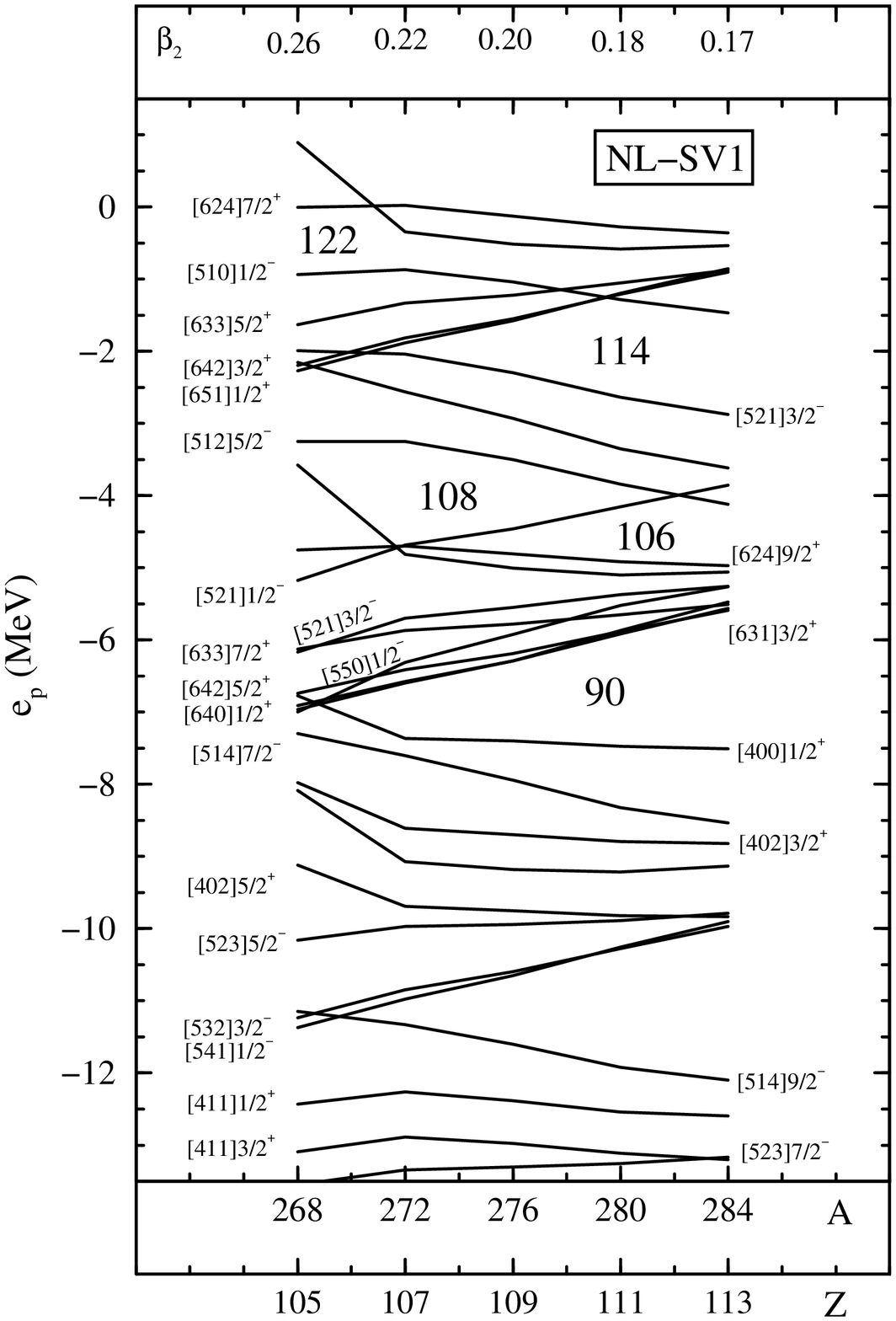}}
}
\caption{The proton Nilsson single-particle levels obtained from the RMF 
calculations with NL-SV1 for low-deformed prolate-shaped nuclei of the 
decay chain of $^{288}$115 are shown.}
\label{fig:9}       
\end{figure}

\begin{figure}
\resizebox{0.50\textwidth}{!}{%
  \rotatebox{0}{\includegraphics{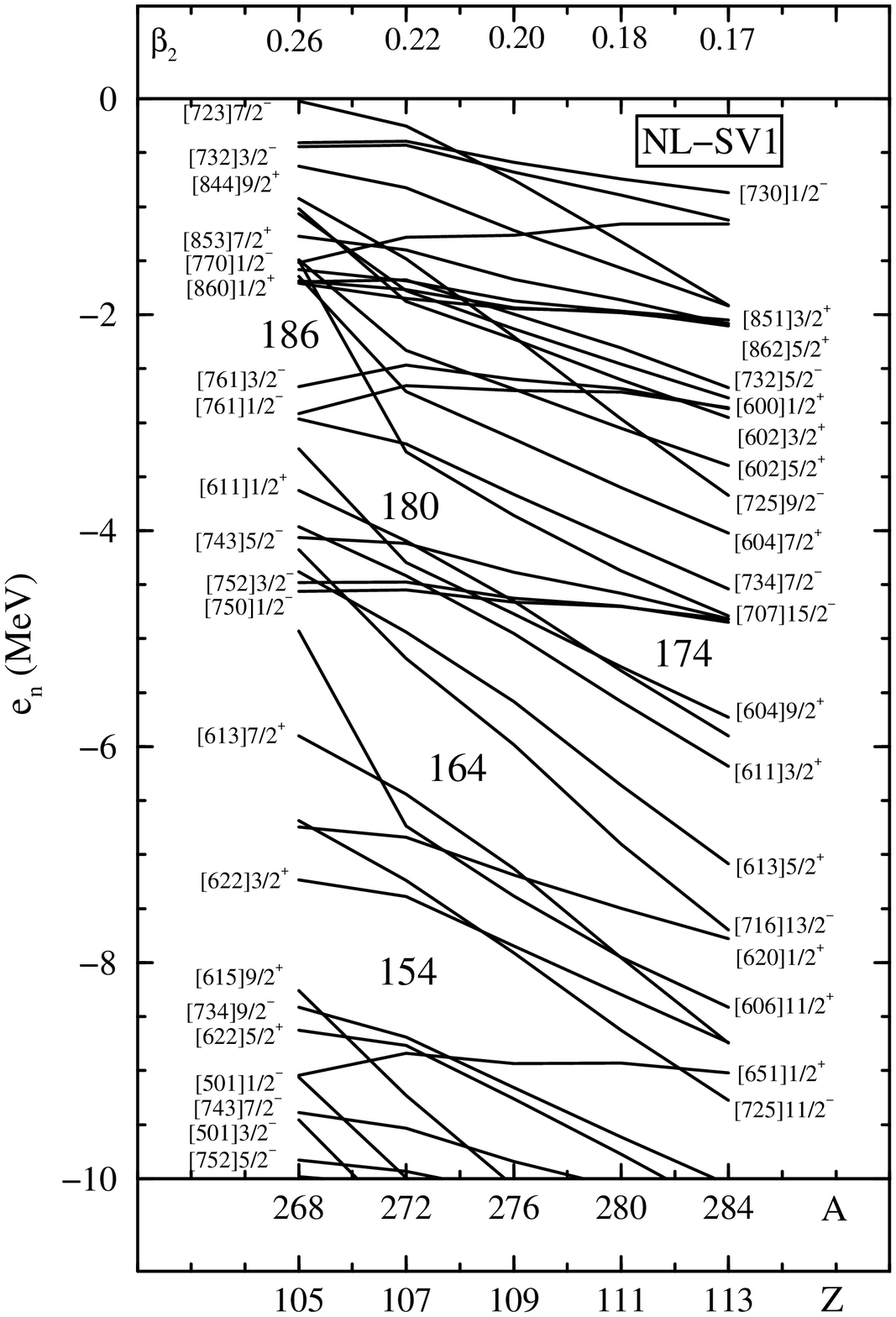}}
}
\caption{The neutron Nilsson single-particle levels obtained from the RMF 
calculations with NL-SV1 for low-deformed prolate-shaped nuclei of the decay
chain of $^{288}$115 are shown.}
\label{fig:10}       
\end{figure}

\subsection{Single-particle levels}

The Nilsson single-particle levels obtained with NL-SV1 for the
nuclei of chain $^{288}$115 are shown in Figs. 9 and 10. 
The proton single-particle levels for nuclei in this chain are 
presented in Fig. 9. The corresponding ground-state
$\beta_2$ values for each nucleus are indicated
in the top panel of the figure. It shows a decrease from 0.26 to 0.17 
in going from $^{268}$Db $(Z=109)$ to $^{284}$113. This decrease 
is accompanied by an arrival
of a deformed shell gap at $Z=114$ as seen in the figure. There are also
minor deformed shell gaps at $Z=106$ and $Z=108$. As the deformation is
changing only slightly and the Fermi energy is showing a marginal
change in going from the left to the right, the level crossing in
Fig. 9 is not such a dominant feature. However, shifting of the deformed 
levels partly due to the change in deformation and partly due to the 
change in the chemical potential, deformed shell gaps are being created
at $Z=108$ for the lower $Z$ nuclei and at $Z=114$ for the higher $Z$ nuclei.
There is, however, no spherical $Z=114$ closed shell as suggested in
models such as FRDM \cite{FRDM.95}.

\begin{figure}
\resizebox{0.50\textwidth}{!}{%
  \rotatebox{0}{\includegraphics{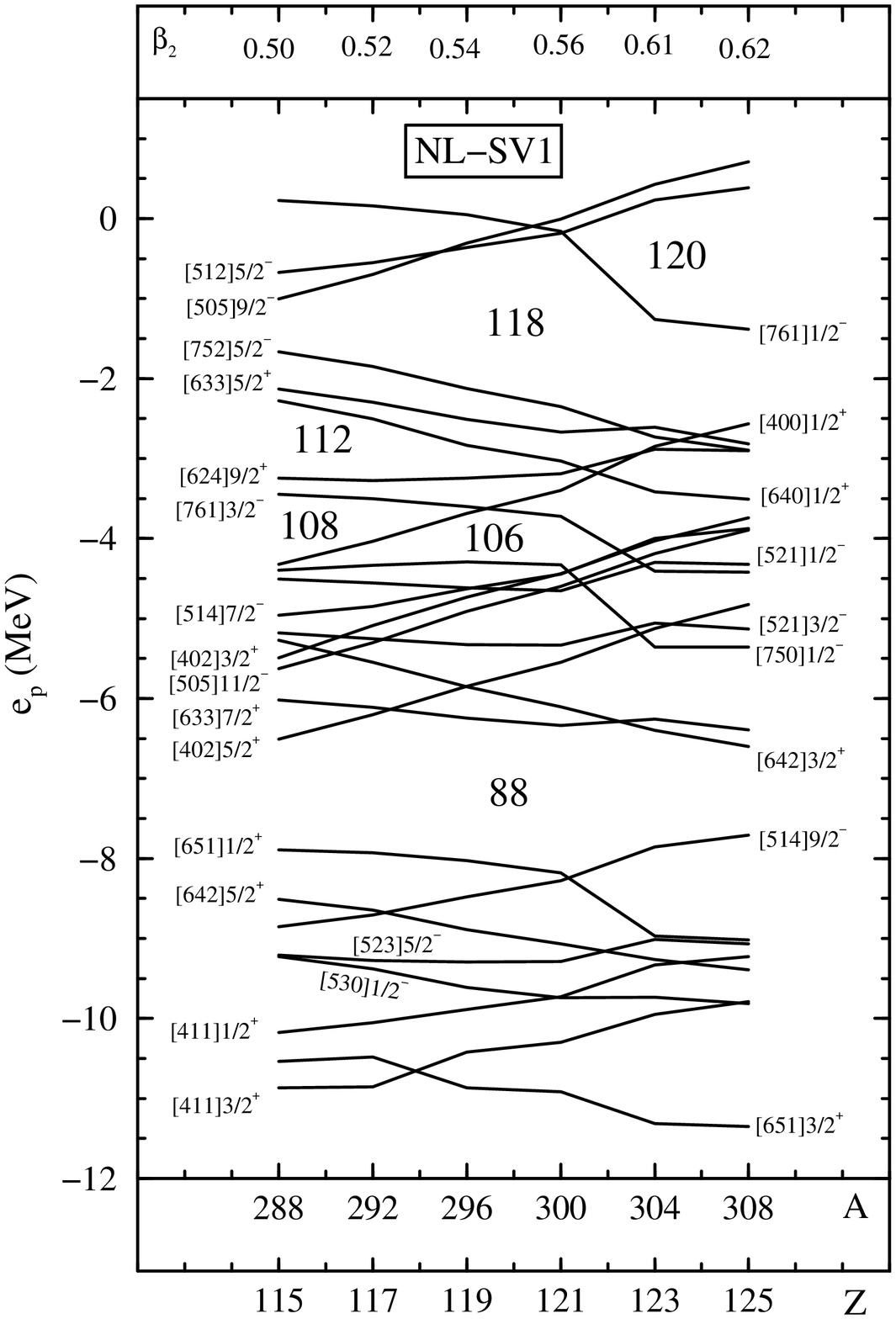}}
}
\caption{The proton Nilsson single-particle levels obtained
from RMF calculations with NL-SV1 for superdeformed prolate  nuclei,
which are $\alpha$-decay precursors of the nucleus $^{288}$115.}
\label{fig:11}       
\end{figure}

\begin{figure}
\resizebox{0.50\textwidth}{!}{%
  \rotatebox{0}{\includegraphics{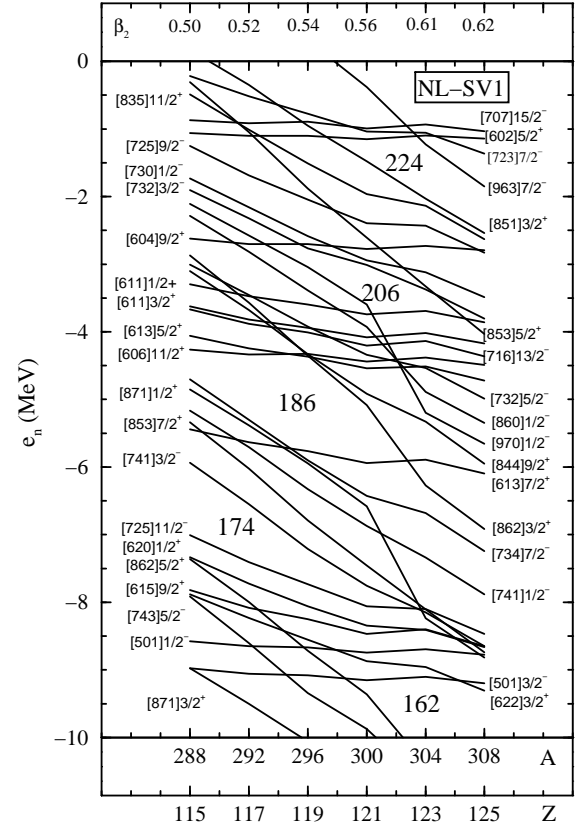}}
}
\caption{The neutron Nilsson single-particle levels obtained
from RMF calculations with NL-SV1 for superdeformed prolate  nuclei,
which are $\alpha$-decay precursors of the nucleus $^{288}$115.}
\label{fig:12}       
\end{figure}

The neutron single-particle levels in Fig. 10 show a 
shell gap at $N=154$ and $N=164$.
The latter is inevitably associated with the nuclei $^{268}$Db and
$^{272}$Bh. For the heavier nuclei with $A=280$ and $A=284$ the deformed
shell gap evolves at $N=174$. This has been observed to the case in
several other models. It can be seen that a frequent
crossing of levels creates a bunching and thus deformed shell gaps
at $N=154$, $N=164$, $N=174$ and $N=180$ are produced.
The well-known shell gap at $N=184$ is not seen in this figure,
though it is not relevant to mass numbers shown in this figure.

The proton Nilsson orbitals for the precursors of the nucleus
$^{288}$115 are shown in Fig. 11.  For these nuclei the ground
state quadrupole deformation $\beta_2$ increases from 0.50 for
$^{288}$115 to 0.62 for $^{308}$125. Because these nuclei are
superdeformed, there is a significant level crossing in this diagram.
A deformed shell gap at $Z=112$ is seen. However, this shell gap
is much less relevant to the higher $Z$ values of nuclei considered
in this figure. For the experimentally observed nucleus
$^{288}$115  \cite{Oganess.04}, there is no perceptible observation of
a deformed shell gap at $Z=114$. This gap which was seen in the
lighter counterparts of the chain is washed out here. In contrast, a
strong shell gap is observed at $Z=118$. This shell gap in the highly deformed
region provides an extra stability to the nuclei concerned. This has already
been seen in terms of lower $Q_\alpha$ values for nuclei with higher
$Z$ as shown in Fig.~5. A possible deformed shell closure at $Z=120$ is also
observed, but for very heavy superheavies. Here the nuclide $^{308}$125 
runs close to the proton drip line.

The corresponding neutron Nilsson orbitals for the superdeformed
superheavy nuclei which are heavier parents of the nucleus $^{288}$115
are shown in Fig. 12.  The neutron number for these nuclei spans the range
$N=173-183$. For the lighter counterparts of the chain one can see a
deformed shell gap at $N=174$, that diminishes in going to the heavier
nuclei. There is another deformed shell gap at $N=186$. However,
we do not see a shell gap at $N=184$ in the deformed region.
Moreover, nuclei in the vicinity of this neutron number are assuming
superdeformed prolate shapes. This discounts the possibility of a
deformed shell gap at $N=184$ for these $Z$ values. For Z values much
lower than those considered in this work, a spherical shell gap at $N=184$
can not be ruled out within this model. Comprehensive calculations
for much of the superdeformed region with this model in future would
clarify the situation.

A lowering in energy of a large number of Nilsson orbitals in going
from $^{288}$115 to $^{300}$121 is accompanied by a slight increase
in the quadrupole deformation. However, splitting of levels
by the strong deformation leads to a significant increase in the
binding energy of nuclei. A sudden jump in the $\beta_2$ value
from the nucleus $^{300}$121 to $^{304}$123 also results in a kink in
several levels and a further lowering in the energy of several Nilsson
orbitals is anticipated. Therefore, this adds to the total
binding energy of nuclei such as $^{304}$123 and $^{308}$125.
The consequence of this has already been seen in Figs. 5 and 6,
where the $Q_\alpha$ value of heavier superheavies is
smaller by $\sim$ 1-2 MeV as compared to FRDM. 
This would inevitably provide a larger stability and
consequently a larger $\alpha$-decay half-lives as compared to the
FRDM. Thus, the present calculations with NL-SV1 present an
optimistic picture for producing superheavy nuclei heavier than
$Z=115$.

\subsection{Shell effects in nuclei and consequences}

As mentioned in Section I, the spin-orbit
interaction and consequently its effect on the shell gaps and
shell effects has a major influence on creation of magic numbers.
It is, however, still not clear as to how the spin-orbit interaction
extrapolates in the extreme limits of the periodic table such
as near the r-process path and for extreme superheavy nuclei.
We believe that both these regions have a bearing upon
each other. Inevitably, how the shell effects would transcend
and extrapolate in unknown regions would be decisive in
r-process nucleosynthesis of heavy nuclei as well as in
synthesis of extreme superheavy nuclei. We put up the case
in Section II that if the shell effects are strong along the
line of stability, these would extrapolate strongly in the extreme
regions and vice-versa. This was demonstrated reasonably well
in ref. \cite{Sharma.02}.

Here a comment on the strength of the shell effects would be in order.
It is well known that the macroscopic-microscopic approaches such
as FRDM \cite{Moeller.97} exhibit shell effects that are stronger
at $N=82$ and $N=132$ in going to extremely rich regions of the r-process
path. The strong shell effects at these magic numbers in FRDM has a
consequence on the r-process nuclear abundances. Using the data from
the FRDM in network chain calculations, there is a considerable
shortfall (troughs) in the abundance of r-process nucleosynthesis
of nuclei below the peaks at $A \sim 130$ and $A \sim 190$ \cite{Kratz.93}.
These peaks in the abundance curve correspond to the above neutron
magic numbers. Therefore, the strong shell effects in FRDM are not
consistent with the r-process nuclear abundances which
require a weakening of the shell effects in going to the
r-process region. The strong nature of the shell effects in the
FRDM is evidently carried away also in the region of superheavy
nuclei. This becomes apparent in going to extreme superheavies in
the neighborhood of $Z=120$ as shown in Fig. 5 and 6,
where large values of $Q_\alpha$ are predicted by the FRDM.

The macroscopic-microscopic calculations of ref. \cite{Muntian.03}
are of the similar origin as the FRDM. These calculations
have been successful in describing superheavy nuclei below $Z=109$
as illustrated in Fig. 2 of ref. \cite{Oganess.04}. As mentioned earlier,
results of ref. \cite{Muntian.03} overestimate the $Q_\alpha$ values
(see Fig. 5 and 6) for superheavy nuclei above $Z=111$ \cite{Oganess.04}.
However, comparing the general trend of the 
predictions from macroscopic-microscopic
calculations of ref. \cite{Muntian.03} with those of the FRDM
near $Z=120$, one finds a reasonably good agreement between the two
model predictions on $Q_\alpha$ values. Both the models predict larger
values for $Q_\alpha$ in the extreme region. This suggests that similar
to the FRDM, strong shell effects are also carried over to the extreme
region in the calculations of ref. \cite{Muntian.03}. This would indeed
lead to very small $\alpha$-decay half-lives for very heavy superheavy
nuclei. On the other hand, the less strong shell effects in microscopic
calculations with NL-SV1 predict $Q_\alpha$ values that are smaller
than those of the FRDM. This shows an immense importance of as to how
the shell effects extrapolate in the extreme regions and how the shell
effects influence properties of nuclei in these regions.

\section{Conclusion}

We have performed relativistic mean-field calculations for the recently
observed alpha-decay chains of nuclei $^{288}$115 and $^{287}$115
using the Lagrangian set NL-SV1 with the vector self-coupling of
$\omega$-meson. The $Q_\alpha$ values have been calculated. It is
shown that across the experimentally observed chains, the NL-SV1
values show a rather smooth behaviour with the change in the Z values of
the nuclei. The $Q_\alpha$ values obtained with NL-SV1 without blocking
of odd-particle pairing describe much of the experimental data very well.
In comparison, the macroscopic-microscopic calculations \cite{Muntian.03}
overestimate the experimental values systematically. The $\alpha$-decay
half-lives calculated with $Q_\alpha$ values from NL-SV1 using the
Viola-Seaborg systematics show a reasonably good agreement with the
experimental values deduced from the recent experiment
\cite{Oganess.04} with the exception of a few nuclei.

The decay products of the both the chains with $Z=$ 107, 109, 111 and 113
are shown to be prolate deformed with a deformation varying from
$\beta_2 \sim 0.22$ for $Z=107$ to $\beta_2 \sim 0.17$ for $Z=113$. Curiously,
the nuclei $^{284}$113 and $^{283}$113 in both the chains exhibit a
shape-coexistence of a low prolate shape with $\beta_2 \sim 0.17$
and a superdeformed prolate shape with $\beta_2 \sim 0.50$. In both
the cases, the low $\beta_2$ shape corresponds to the lowest energy
minimum. The parent nuclei $^{288}$115 and $^{287}$115 of both the
chains, however, exhibit a very large deformation akin to a
superdeformed shape. 

We have also made predictions for heavier elements in the two chains
of $^{288}$115 and $^{287}$115. It is shown that with NL-SV1 the
ground-state of the higher $Z$ counterparts of both the chains are
superdeformed with $\beta_2$ values in the vicinity of 0.60 with a
tendency of an increase in $\beta_2$ value in going to heavier elements.
This can be attributed to a softening of the shell effects in going to
the extreme case. The $Q_\alpha$ values obtained with NL-SV1 for nuclei
with $Z=$ 117, 119, 121, 123 and 125 are about 1-2 MeV
lower than the predictions of the FRDM \cite{FRDM.95}.
This has a consequence that $\alpha$-decay half-lives of the heavier
nuclei are 2-3 orders of magnitude larger than the existing
predictions of very short half-lives from the macroscopic-microscopic
models \cite{Muntian.03,Moeller.97}. This scenario would put synthesis
of heavier superheavy elements on a feasible footing.

\begin{acknowledgments}
This work is supported by the Research Administration Project No.
SP05/02 of Kuwait University. We thank Professor Y.K. Gambhir
for useful comments and discussion.
\end{acknowledgments}


\end{document}